\documentclass[9pt]{article}
\usepackage[preprint]{spconf}
\usepackage{amsmath,amstext}
\usepackage[T1]{fontenc}
\usepackage{amssymb}
\usepackage{graphicx}
\usepackage{cite}
\usepackage{caption}
\copyrightnotice{\copyright2020 IEEE}
\toappear{To appear at ICASSP 2020}

\title{BINAURAL AUDIO SOURCE REMIXING WITH \\ MICROPHONE ARRAY LISTENING DEVICES}
\name{Ryan M. Corey and Andrew C. Singer\thanks{This material is based upon work supported by the National Science Foundation under Grant No. 1919257.}}
\address{University of Illinois at Urbana-Champaign}
\ninept

\begin{document}
\maketitle
\begin{abstract}
\small
Augmented listening devices, such as hearing aids and augmented reality
headsets, enhance human perception by changing the sounds that we hear. 
Microphone arrays can improve the performance of listening
systems in noisy environments, but most array-based listening systems
are designed to isolate a single sound source from a mixture. This
work considers a source-remixing filter that alters the relative level
of each source independently. Remixing rather than separating sounds
can help to improve perceptual transparency: it causes less distortion
to the signal spectrum and especially to the interaural cues that
humans use to localize sounds in space.
\end{abstract}


\section{Introduction}

Audio signal processing can be used to change the way that humans
experience the world around them. Augmented listening (AL) devices,
such as hearing aids and augmented reality headsets \cite{valimaki2015assisted},
enhance human hearing by altering the sounds presented to a listener.
One of the most important functions of listening devices is to help
humans to hear better in noisy environments with many competing sound
sources. Unfortunately, many listening devices today perform poorly
in noisy environments where users need them most.

One reliable way to reduce unwanted noise and improve intelligibility
in noisy situations is with microphone array processing. Arrays of
several spatially separated microphones can be used to perform beamforming
and isolate sounds from a particular direction \cite{gannot2017consolidated}.
Beamformers are widely used in machine listening systems, for example
for automatic speech recognition, to preprocess a sound source of
interest. For decades, researchers have tried to incorporate similar
beamforming technology into human listening devices \cite{soede1993assessment,Greenberg2001survey,Doclo2008,doclo2015magazine}.
Most past studies have taken the same approach used in machine listening:
steering a beam to isolate a single sound source of interest and attenuate
all others. Although this approach has been shown to improve intelligibility
in noise \cite{soede1993assessment,Greenberg2001survey},
listening devices using large arrays have never been commercially
successful.

The human auditory system is different from a machine listening algorithm.
Isolating a single sound source, while it might improve intelligibility,
would seem unnatural to the listener and it could interfere with the
auditory system's natural source separation and scene analysis capabilities.
Humans rely on spectral and temporal patterns as well as spatial cues, such as interaural time and level differences,
to distinguish between sound sources and remain aware of their environment
\cite{bregman1994auditory,bronkhorst2000cocktail}. Single-target beamformers can
distort the spectral and spatial cues of all non-target sound sources \cite{doclo2006theoretical}.
Beamformers can be designed to preserve the interaural cues of multiple
sources by deliberately including components of non-target sources in the
filter output \cite{vandenbogaert2008thesis,marquardt2016development,koutrouvelis2018multi}.
Early binaural beamformers simply added a fraction of the unprocessed
signal into the output \cite{klasen2007binaural,van2009speech,cornelis2010theoretical},
while more recent methods apply explicit constraints to the filter's
response to those signals \cite{marquardt2015theoretical,hadad2016binaural}.
It is also possible to constrain only the interaural cues and not
the spectral distortion of the sources \cite{hadad2015theoretical,koutrouvelis2016improved}.
In general, the more the background sources are preserved, the less
their interaural cues are distorted \cite{cornelis2010theoretical}.

\begin{figure}
\begin{centering}
\includegraphics{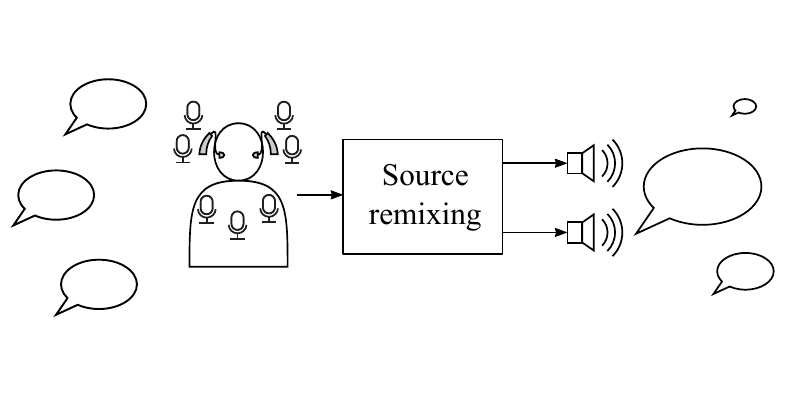}
\par\end{centering}
\caption{\label{fig:motivation}The proposed source-remixing system uses a
microphone array and space-time filters to apply separate processing
to many sound sources, altering the auditory perception of the listener.}
\end{figure}

In this work, we approach augmented listening as a \emph{source remixing}
problem. Rather than trying to isolate a single sound source, we use a microphone array to apply different
processing to many sound sources, as shown in Fig. \ref{fig:motivation}.
The system is analogous to the mixing process in a music or film studio,
where each instrument, dialogue track, and sound effect in the mixture
is individually adjusted to provide a pleasant listening experience. The
processing applied to each source depends on the type of sound, the
acoustic environment, and the listener's preferences. For example,
a normal-hearing user in a quiet room might prefer no processing at
all, while a hearing-impaired user in a noisy restaurant could use
aggressive noise reduction similar to single-target beamforming. A
similar remixing approach has been proposed for television broadcasts
with object-based audio encodings: listeners can tune the relative
levels of sources to trade off between immersiveness and intelligibility
\cite{jot2015dialog,shirley2017personalized}. Using a powerful microphone
array, we could allow listeners to perform that same tuning for real-world
sounds.

In this work, we consider the choice of relative levels of sound sources
in a source-remixing filter. Further perceptual research
and clinical studies will be needed to understand how to select these
levels to optimize the listening experience for different AL applications.
In the meantime, we can characterize the engineering tradeoffs
of such a system. Intuitively, the less we try to separate
the sound sources---that is, the more transparent the listening
experience---the easier it should be to implement the remixing system
and the more natural it should sound to the listener. Studies of musical
source separation have shown that remixing can reduce unpleasant artifacts
and distortion compared to complete separation \cite{wierstorf2017perceptual}.
Applying independent processing to different sound sources can also
reduce the distortion introduced by nonlinear processing algorithms
such as dynamic range compression \cite{corey2017compression}. Here
we focus on the benefits of remixing for spectral and spatial distortion
in a linear time-invariant space-time filter. We show that the less
the mixture is altered, the lower the resulting distortion.

\section{Space-time filter for source remixing}

\subsection{Signal model}

Consider an array of $M$ microphones. By convention, it is assumed
that microphone 1 is in or near the left ear and microphone 2 is in
or near the right ear; these act as reference microphones for the
purposes of preserving interaural cues. The mixture is assumed to
include $N$ \emph{source channels.} A source channel is defined as
a set of sounds that are to be processed as a group, that is, for
which the desired response of the system is the same. A source channel
could be an individual talker, a group of talkers, diffuse noise,
or all sounds of a certain type, for example. 

Let $\mathbf{x}(t)\in\mathbb{R}^{M}$ be the vector of continuous-time
signals observed by the microphones. It is the sum of $N$ source
images $\mathbf{c}_{n}(t)$ due to the source channels:
\begin{equation}
\mathbf{x}(t)=\sum\nolimits_{n=1}^{N}\mathbf{c}_{n}(t).
\end{equation}
Let $\mathbf{y}(t)\in\mathbb{R}^{2}$ be the desired output of the
system. By convention, channel 1 is output to the left ear and channel
2 is output to the right ear. For the purposes of this work, assume
that the desired processing to be applied to each source channel $n$ is
linear and time-invariant so that it can be characterized by an impulse
response matrix $\mathbf{g}_{n}(\tau)\in\mathbb{R}^{2\times M}$. The desired output is therefore
\begin{equation}
\mathbf{y}(t)=\sum_{n=1}^{N}\underbrace{\int_{-\infty}^{\infty}\mathbf{g}_{n}(\tau)\mathbf{c}_{n}(t-\tau)\,\mathrm{d}\tau}_{\mathbf{d}_{n}(t)}.
\end{equation}
The signal $\mathbf{d}_{n}(t)$ is the desired output image for source
channel $n$.

In a perceptually transparent listening system, the listener should
perceive $\mathbf{d}_{n}(t)$ as a natural sound that replaces $\mathbf{c}_{n}(t)$.
To ensure that $\mathbf{d}_{n}(t)$ sounds natural, it should have
a similar frequency spectrum to $\mathbf{c}_{n}(t)$---or a deliberately
altered spectrum, for example to amplify high frequencies for hearing-impaired
listeners---and it should be perceived as coming from the same direction.
It should also have imperceptibly low delay, which limits the amount
of frequency-selective processing that can be applied \cite{corey2018delay}.

Because this work is designed to highlight interaural cue preservation,
we restrict our attention to desired responses of the form 
\begin{equation}
\mathbf{g}_{n}(\tau)=g_{n}(\tau)\left[\begin{matrix}\mathbf{e}_{1} & \mathbf{e}_{2}\end{matrix}\right]^{T},\label{eq:desired_binaural}
\end{equation}
where $\mathbf{e}_{m}$ is the unit vector with value 1 in position
$m$ and 0 elsewhere. In other words, the desired output in the left (respectively
right) ear is the source signal as observed by the left (respectively
right) reference microphone and processed by a diotic impulse response
$g_{n}(\tau)$. If this desired response could be achieved exactly,
the same processing would be applied to the signals in both ears and
the interaural cues of all source channels would be perfectly preserved.

\subsection{Space-time filtering}

The output $\hat{\mathbf{y}}(t)\in\mathbb{R}^{2}$ of the system is
produced by a linear time-invariant space-time filter $\mathbf{w}(\tau)\in\mathbb{R}^{2\times M}$
such that
\begin{align}
\mathbf{\hat{y}}(t) & =\int_{-\infty}^{\infty}\mathbf{w}(\tau)\mathbf{x}(t-\tau)\,\mathrm{d}\tau\\
 & =\sum_{n=1}^{N}\underbrace{\int_{-\infty}^{\infty}\mathbf{w}(\tau)\mathbf{c}_{n}(t-\tau)\,\mathrm{d\tau}}_{\hat{\mathbf{d}}_{n}(t)}.
\end{align}
The signal $\hat{\mathbf{d}}_{n}(t)\in\mathbb{R}^{2}$ is the output
image for source channel $n$; it is the processed version of the
original source image $\mathbf{c}_{n}(t)$ perceived by the listener.

In an AL system, sound signals must be processed in real time with
a total delay of no more than a few milliseconds to avoid disturbing
distortion. Thus, $\mathbf{w}(\tau)$ should be a causal filter that
predicts a possibly delayed version of $\mathbf{y}(t)$. It has been
shown that such a constraint limits the performance of a space-time
filter, especially for small arrays \cite{corey2018delay}. Because space-time filters are most conveniently studied in the frequency domain, $\mathbf{w}(\tau)$ is allowed to be noncausal in our mathematical analysis, but the discrete-time filters implemented in Sec. \ref{sec:experiments} are causal with realistic delay constraints.

\subsection{Weighted remixing filter}

We would like to choose $\mathbf{w}(\tau)$ so that $\hat{\mathbf{d}}_{n}(t)\approx\mathbf{d}_{n}(t)$
for all $n$. To make the space-time filter as flexible
as possible, we will use the multiple-speech-distortion-weighted multichannel
Wiener filter (MSDW-MWF) \cite{markovich2012weighted}. The MSDW-MWF
is a generalization of the well-known speech-distortion-weighted multichannel
Wiener filter (SDW-MWF) \cite{Doclo2005}, which allows the system
designer to trade noise for spectral distortion of a single target
source. The MSDW-MWF minimizes the following weighted squared-error
cost function:
\begin{equation}
\mathrm{Cost}=\sum_{n=1}^{N}\lambda_{n}\mathbb{E}\left[\left\Vert \hat{\mathbf{d}}_{n}(t)-\mathbf{d}_{n}(t)\right\Vert ^{2}\right],
\end{equation}
where $\mathbb{E}$ denotes statistical expectation and $\lambda_{n}$
is a \emph{distortion weight} that controls the relative importance
of each sound source.

If the source images are statistically uncorrelated with each other,
then the noncausal MSDW-MWF is given by the frequency-domain expression
\begin{equation}
\mathbf{W}(\Omega)=\left(\sum_{n=1}^{N}\lambda_{n}\mathbf{G}_{n}(\Omega)\mathbf{R}_{\mathbf{c}_{n}}(\Omega)\right)\underbrace{\left(\sum_{n=1}^{N}\lambda_{n}\mathbf{R}_{\mathbf{c}_{n}}(\Omega)\right)}_{\bar{\mathbf{R}}_{\mathbf{x}}(\Omega)}\biggr.^{-1},
\end{equation}
where $\mathbf{G}_n(\Omega)$ is the Fourier transform of $\mathbf{g}_n(\tau)$ and $\mathbf{R}_{\mathbf{c}_{n}}(\Omega)$ is the power spectral
density of source image $\mathbf{c}_{n}(t)$ for $n=1,\dots,N$, which must be measured or estimated. It is assumed that $\bar{\mathbf{R}}_{\mathbf{x}}(\Omega)$ is invertible, which can be achieved by including a diffuse noise channel with full-rank spectral density. Many
commonly used space-time filters can be derived
from the MSDW-MWF: the multichannel Wiener filter, which minimizes
mean squared error, is a special case with $\lambda_{n}=1$ for all
$n$. Linearly constrained filters such as the minimum variance distortionless
response beamformer are limiting cases as the distortion weights approach
infinity \cite{gannot2017consolidated}.

The remainder of the analysis will be performed in the frequency domain
and we omit the frequency variable $\text{\ensuremath{\Omega}}$ for
brevity.

\subsection{Spectral distortion of the remixing filter}

The following identity will be useful in analyzing the performance
of the MSDW-MWF:
\begin{align}
\mathbf{W} & =\sum_{m=1}^{N}\lambda_{m}\mathbf{G}_{m}\mathbf{R}_{\mathbf{c}_{m}}\bar{\mathbf{R}}_{\mathbf{x}}^{-1}\\
 & =\mathbf{G}_{n}+\sum_{m=1}^{N}\lambda_{m}(\mathbf{G}_{m}-\mathbf{G}_{n})\mathbf{R}_{\mathbf{c}_{m}}\bar{\mathbf{R}}_{\mathbf{x}}^{-1}.\label{eq:msdw-mwf-identity}
\end{align}
If the filter is computed using the true source statistics, then the
weighted error spectral density of the MSDW-MWF is given by
\begin{align}
\bar{\mathbf{R}}_{\mathrm{err}} & =\sum_{n=1}^{N}\lambda_{n}\left(\mathbf{G}_{n}-\mathbf{W}\right)\mathbf{R}_{\mathbf{c}_{n}}\left(\mathbf{G}_{n}-\mathbf{W}\right)^{H}\\
 & =\sum_{n=1}^{N}\sum_{m=1}^{M}\lambda_{n}\lambda_{m}\left(\mathbf{G}_{n}-\mathbf{G}_{m}\right)\mathbf{R}_{\mathbf{c}_{m}}\bar{\mathbf{R}}_{\mathbf{x}}^{-1}\mathbf{R}_{\mathbf{c}_{n}}\left(\mathbf{G}_{n}-\mathbf{W}\right)^{H}\\
 & =\sum_{n=1}^{N}\sum_{m=1}^{M}\lambda_{n}\lambda_{m}\left(\mathbf{G}_{n}-\mathbf{G}_{m}\right)\mathbf{R}_{\mathbf{c}_{m}}\bar{\mathbf{R}}_{\mathbf{x}}^{-1}\mathbf{R}_{\mathbf{c}_{n}}\mathbf{G}_{n}^{H},\label{eq:weighted_err}
\end{align}
where the last step follows from the orthogonality principle. Thus,
the performance of the remixing filter can be expressed in terms of
pairs of source channels. It is clear from this expression that if
$\mathbf{G}_{n}=\mathbf{G}_{m}$, then the $(m,n)$ source pair contributes
nothing to the error of the system: if we wish to apply the same processing
to both sources, then we need not separate them.

If each desired response has the form of (\ref{eq:desired_binaural})
and if the Fourier transform $G_{n}(\Omega)$ of $g_{n}(\tau)$ is
real-valued for all $n=1,\dots,N$, then by manipulating (\ref{eq:weighted_err})
the weighted error spectra in the left and right ears can be written
\begin{align}
\bar{R}_{\mathrm{err}}^{\mathrm{left}} & =\frac{1}{2}\sum_{n=1}^{N}\sum_{m=1}^{M}\lambda_{n}\lambda_{m}\left|G_{n}-G_{m}\right|^{2}\mathbf{e}_{1}^{T}\mathbf{R}_{\mathbf{c}_{m}}\bar{\mathbf{R}}_{\mathbf{x}}^{-1}\mathbf{R}_{\mathbf{c}_{n}}\mathbf{e}_{1}\\
\bar{R}_{\mathrm{err}}^{\mathrm{right}} & =\frac{1}{2}\sum_{n=1}^{N}\sum_{m=1}^{M}\lambda_{n}\lambda_{m}\left|G_{n}-G_{m}\right|^{2}\mathbf{e}_{2}^{T}\mathbf{R}_{\mathbf{c}_{m}}\bar{\mathbf{R}}_{\mathbf{x}}^{-1}\mathbf{R}_{\mathbf{c}_{n}}\mathbf{e}_{2}.
\end{align}
Thus, the performance of the remixing filter depends on the differences
between desired responses and on the spatial and spectral separability of the
signal pairs. 

\section{Distortion of interaural cues}

The interaural level difference (ILD) and interaural phase difference
(IPD) can both be derived from the interaural transfer function (ITF).
If the source image $\mathbf{c}_{n}(t)$ for channel $n$ has a Fourier
transform $\mathbf{C}_{n}(\Omega)$, then the input and output ITFs
for source channel $n$ are
\begin{align}
\mathrm{ITF}_{n}^{\mathrm{in}} & =\frac{\mathbf{e}_{2}^{T}\mathbf{C}_{n}}{\mathbf{e}_{1}^{T}\mathbf{C}_{n}}\\
\mathrm{ITF}_{n}^{\mathrm{out}} & =\frac{\mathbf{e}_{2}^{T}\hat{\mathbf{D}}_{n}}{\mathbf{e}_{1}^{T}\hat{\mathbf{D}}_{n}}=\frac{\mathbf{e}_{2}^{T}\mathbf{W}\mathbf{C}_{n}}{\mathbf{e}_{1}^{T}\mathbf{W}\mathbf{C}_{n}}.
\end{align}
The ILD and IPD are the magnitude and phase of the ITF:
\begin{equation}
\mathrm{ILD}_{n}=20\log_{10}|\mathrm{ITF}_{n}|\quad\text{and}\quad\mathrm{IPD}_{n}=\angle\mathrm{ITF}_{n}.
\end{equation}

\subsection{Spatial distortion of the MSDW-MWF}

The output ITF for the MSDW-MWF (\ref{eq:msdw-mwf-identity}) with
binaurally matched responses (\ref{eq:desired_binaural}) is 
\begin{equation}
\mathrm{ITF}_{n}^{\mathrm{out}}=\frac{G_{n}\mathbf{e}_{2}^{T}\mathbf{C}_{n}\!+\!\sum_{m=1}^{N}\!\lambda_{m}\left(G_{m}\!-\!G_{n}\right)\mathbf{e}_{2}^{T}\mathbf{R}_{\mathbf{c}_{m}}\bar{\mathbf{R}}_{\mathbf{x}}^{-1}\mathbf{C}_{n}}{G_{n}\mathbf{e}_{1}^{T}\mathbf{C}_{n}\!+\!\sum_{m=1}^{N}\!\lambda_{m}\left(G_{m}\!-\!G_{n}\right)\mathbf{e}_{1}^{T}\mathbf{R}_{\mathbf{c}_{m}}\bar{\mathbf{R}}_{\mathbf{x}}^{-1}\mathbf{C}_{n}}
\end{equation}
for $n=1,\dots,N$. Notice that if the second terms in the numerator
and denominator were removed, the output ITF would be identical to
the input ITF. This would be the case if the same processing
were applied to every source channel so that all $G_{m}-G_{n}=0$
or if the sources were fully separable so that $\mathbf{R}_{\mathbf{c}_{m}}\bar{\mathbf{R}}_{\mathbf{x}}^{-1}\mathbf{C}_{n}=\mathbf{0}$ for $m\ne n$.

The error in the ILD and IPD are the real and imaginary parts, respectively,
of the logarithm of $\mathrm{ITF}_{n}^{\mathrm{out}}/\mathrm{ITF}_{n}^{\mathrm{in}}$.
If $G_{n}$ and $\mathrm{ITF}_{n}^{\mathrm{in}}$ are nonzero, then
the ITF error for source channel $n$ can be written
\begin{equation}
\mathrm{\Delta ITF}_{n}=\ln\frac{1+\sum_{m=1}^{N}\lambda_{m}\frac{G_{m}-G_{n}}{G_{n}}\frac{\mathbf{e}_{2}^{T}\mathbf{R}_{\mathbf{c}_{m}}\bar{\mathbf{R}}_{\mathbf{x}}^{-1}\mathbf{C}_{n}}{\mathbf{e}_{2}^{T}\mathbf{C}_{n}}}{1+\sum_{m=1}^{N}\lambda_{m}\frac{G_{m}-G_{n}}{G_{n}}\frac{\mathbf{e}_{1}^{T}\mathbf{R}_{\mathbf{c}_{m}}\bar{\mathbf{R}}_{\mathbf{x}}^{-1}\mathbf{C}_{n}}{\mathbf{e}_{1}^{T}\mathbf{C}_{n}}}.\label{eq:itf_err}
\end{equation}

\subsection{First-order approximation}

\begin{figure}
\begin{centering}
\includegraphics[height=3cm]{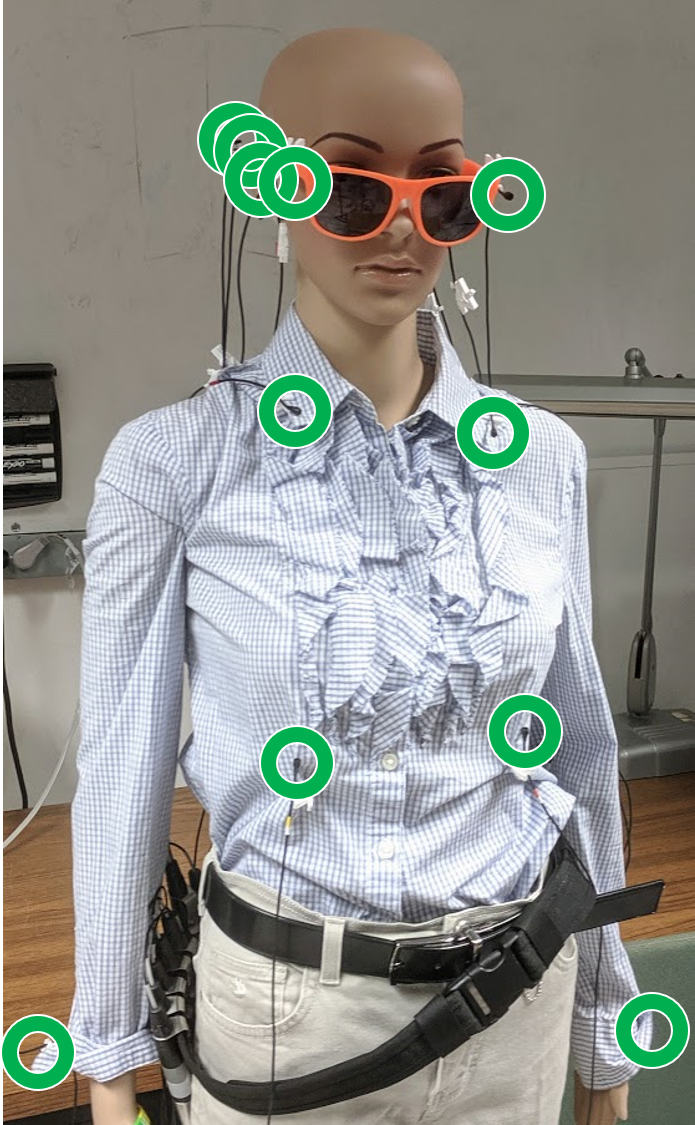}\hfill{}\includegraphics{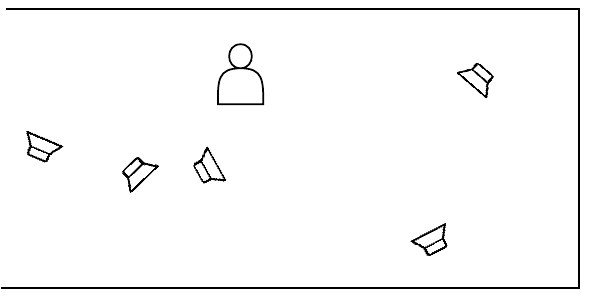}
\par\end{centering}
\caption{\label{fig:layout}Five loudspeakers were placed around an acoustically
treated laboratory. A wearable array includes up to 16 microphones.}
\end{figure}

Using the first-order approximation $\ln\left(1+u\right)\approx u$
for both the numerator and denominator of (\ref{eq:itf_err}), the
logarithmic ITF error is 
\begin{align}
\Delta\mathrm{ITF}_{n} & \approx \nonumber \\ 
& \sum_{m=1}^{N}\!\lambda_{m}\frac{G_{m}\!-\!G_{n}}{G_{n}}\left(\frac{\mathbf{e}_{2}^{T}\mathbf{R}_{\mathbf{c}_{m}}\bar{\mathbf{R}}_{\mathbf{x}}^{-1}\mathbf{C}_{n}}{\mathbf{e}_{2}^{T}\mathbf{C}_{n}}\!-\!\frac{\mathbf{e}_{1}^{T}\mathbf{R}_{\mathbf{c}_{m}}\bar{\mathbf{R}}_{\mathbf{x}}^{-1}\mathbf{C}_{n}}{\mathbf{e}_{1}^{T}\mathbf{C}_{n}}\right).
\end{align}

Furthermore, if diffuse noise were negligible and every source channel were well modeled by a rank-1
spectral density matrix $\mathbf{R}_{\mathbf{c}_{n}}\approx R_{s_{n}}\mathbf{A}_{n}\mathbf{A}_{n}^{H}$
with $\mathbf{C}_{n}$ parallel to the acoustic transfer function
$\mathbf{A}_{n}$ for $n=1,\dots,N$, then the ITF error would be
\begin{align}
\Delta\mathrm{ITF}_{n} & \approx \nonumber \\ 
& \sum_{m=1}^{N}\!\lambda_{m}R_{s_{m}}\frac{G_{m}\!-\!G_{n}}{G_{n}}\mathbf{A}_{m}^{H}\bar{\mathbf{R}}_{\mathbf{x}}^{-1}\mathbf{A}_{n}\left(\frac{\mathbf{e}_{2}^{T}\mathbf{A}_{m}}{\mathbf{e}_{2}^{T}\mathbf{A}_{n}}\!-\!\frac{\mathbf{e}_{1}^{T}\mathbf{A}_{m}}{\mathbf{e}_{1}^{T}\mathbf{A}_{n}}\right)
\end{align}
Thus, spatial distortion depends on the power and distortion weight
of each interfering source, the relative difference in desired responses
between sources, the spatial separability of the sources, and the
difference in interaural cues between source channels. Distant sound
sources that have different acoustic transfer functions will be easier
to separate (small $\mathbf{A}_{m}^{H}\bar{\mathbf{R}}_{\mathbf{x}}^{-1}\mathbf{A}_{n}$)
but also have more different interaural cues (large $\frac{\mathbf{e}_{2}^{T}\mathbf{A}_{m}}{\mathbf{e}_{2}^{T}\mathbf{A}_{n}}-\frac{\mathbf{e}_{1}^{T}\mathbf{A}_{m}}{\mathbf{e}_{1}^{T}\mathbf{A}_{n}}$).
Meanwhile, if $\mathbf{A}_{m}$ is parallel to $\mathbf{A}_{n}$,
the source channels have the same interaural cues and therefore the
source pair does not introduce spatial distortion.

\section{Experiments with a wearable array}
\label{sec:experiments}

\begin{figure}
\begin{centering}
\includegraphics{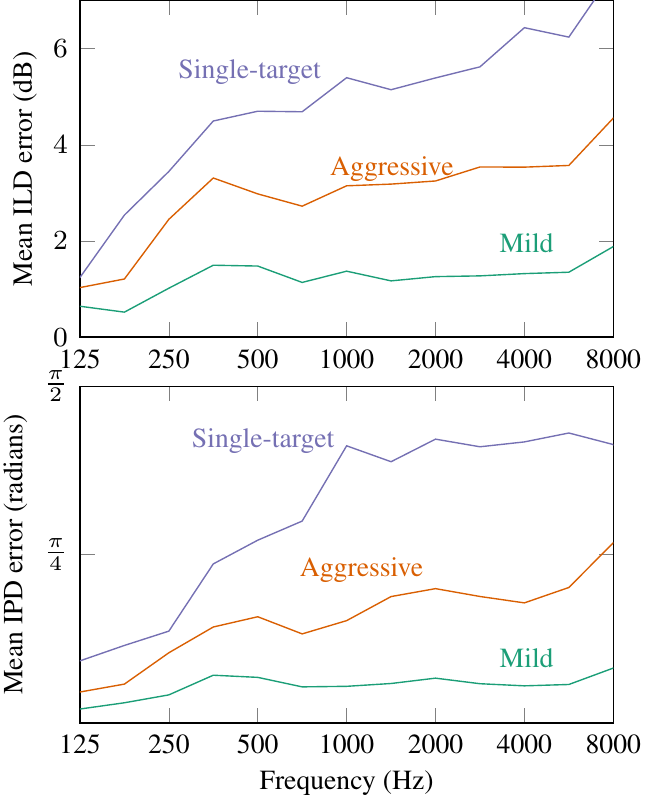}
\par\end{centering}
\caption{\label{fig:binaural_gains}Interaural cue preservation for a binaural
source-remixing filter with 4 microphones and different desired responses.}
\end{figure}

To evaluate the performance of the source-remixing system for augmented
listening applications, it was applied to a wearable microphone array
in a challenging noisy environment. Sixteen microphones were spread
across the body of a mannequin, as shown in Fig. \ref{fig:layout}.
One microphone was placed just outside each ear canal. Although this mannequin does not have realistic
head-related transfer functions, it is suitable for this experiment
because its head has contralateral attenuation that is only slightly
weaker than that of a human \cite{corey2019brtf}. The sound
sources were five speech clips derived from the CSTR VCTK corpus \cite{Veaux2017}
and played through loudspeakers as well as the diffuse mechanical and ventilation noise in the room.

Impulse response measurements and long-term average speech and noise spectra were used to design causal
discrete-time MSDW-MWFs with a delay of 16 ms and a unit pulse response
length of 256 ms. The experimental ITFs of the five speech sources
were measured in the STFT domain using their sample cross-correlations
\cite{cornelis2010theoretical}:
\begin{align}
\mathrm{ITF}_{n}^{\mathrm{in}}[f] & =\frac{\sum_{k}\mathbf{e}_{1}^{T}\mathbf{C}_{\mathrm{stft},n}[k,f]\mathbf{C}_{\mathrm{stft},n}^{H}[k,f]\mathbf{e}_{2}}{\sum_{k}\mathbf{e}_{1}^{T}\mathbf{C}_{\mathrm{stft},n}[k,f]\mathbf{C}_{\mathrm{stft},n}^{H}[k,f]\mathbf{e}_{1}}\\
\mathrm{ITF}_{n}^{\mathrm{out}}[f] & =\frac{\sum_{k}\mathbf{e}_{1}^{T}\hat{\mathbf{D}}_{\mathrm{stft},n}[k,f]\hat{\mathbf{D}}_{\mathrm{stft},n}^{H}[k,f]\mathbf{e}_{2}}{\sum_{k}\mathbf{e}_{1}^{T}\hat{\mathbf{D}}_{\mathrm{stft},n}[k,f]\hat{\mathbf{D}}_{\mathrm{stft},n}^{H}[k,f]\mathbf{e}_{1}},
\end{align}
for $n=1,\dots,5$. The ILD and IPD errors were computed from the
ITFs and averaged over the five directional sources.

Figure \ref{fig:binaural_gains} shows the performance of earpieces
with $M=4$ total microphones for three sets of target responses:
\begin{enumerate}
\item Mild remixing with gains 1.0, 0.8, 0.7, 0.6, and 0.5 on the
speech channels and 20 dB attenuation on the noise channel,
\item Aggressive remixing with gains 1.0, 0.4, 0.3, 0.2, and 0.1
on the speech channels and 20 dB attenuation on noise,
and
\item Single-target beamforming with $G_{1}=1$ and $G_{2}=\cdots=G_{6}=0$.
\end{enumerate}
The mild filter has low interaural cue distortion, the
single-target beamformer severely distorts the non-target sources,
and the aggressive filter falls in between. Distortion is 
mild below a few hundred hertz; these wavelengths are
much larger than a human head and so the ILD and IPD of all sources
are close to zero. 

\begin{figure}
\begin{centering}
\includegraphics{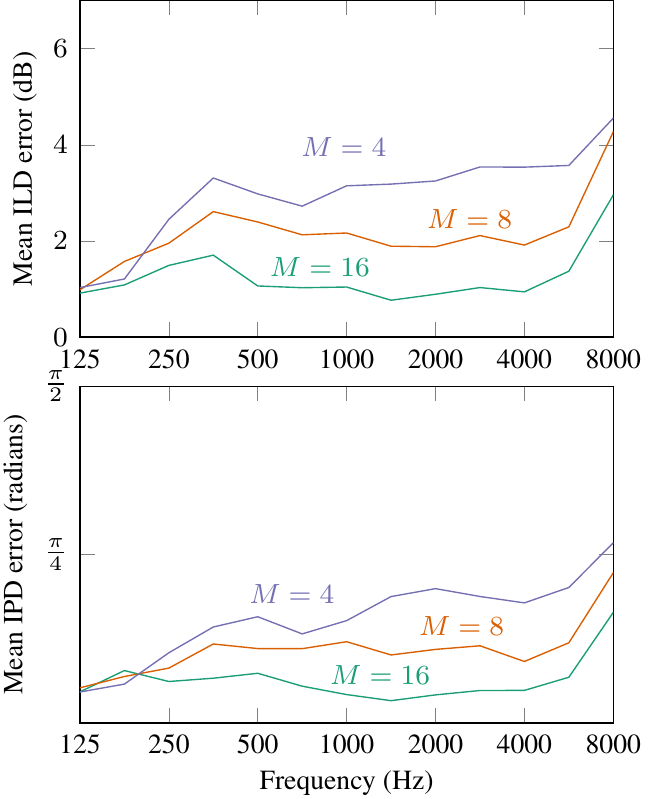}
\par\end{centering}
\caption{\label{fig:binaural_arrays}Interaural cue preservation for a binaural
source-remixing filter with different array configurations.}
\end{figure}

A larger wearable array should be able to apply complex remixing to
more sources than a small earpiece-based array can. Figure \ref{fig:binaural_arrays}
shows the ILD and ITD distortion for the ``aggressive'' remixing
responses with arrays of different sizes. The four-microphone earpiece
array does not have enough degrees of freedom to preserve the interaural
cues of all five directional sources. The 8-microphone head-mounted
array does better, and the 16-microphone upper-body array produces
little distortion in any of the source channels. 

\section{Conclusions}

The design of source-remixing filters requires a tradeoff between
audio enhancement---removing and altering different sound sources
to improve intelligibility---and perceptual transparency. Filters
that alter the signal less, that is, that apply similar desired responses
to the different source channels, cause less spectral distortion and
less interaural cue distortion. They also sound more immersive and
natural to the listener. However, they do not provide as much benefit
in complex noise. Space-time remixing filters with large wearable
microphone arrays could provide the advantages of both approaches:
they have enough spatial resolution to meaningfully suppress strong
background noise, but they have enough degrees of freedom to ensure
that those attenuated noise sources sound natural. 

A full understanding of source-remixing filter design tradeoffs will
require new clinical research. The choice of desired responses will
likely depend on the nature of the sources and environment and on
the preferences of the individual. Compared to conventional beamforming
and source separation, audio source remixing provides a more versatile
approach to human sensory augmentation that could dramatically---but
seamlessly---change how we perceive our world.

\bibliographystyle{ieeetr}
\small \bibliography{../master_references}

\end{document}